**Assess the impacts of human mobility change on COVID-19 dynamics in Arizona, U.S.: a modeling study incorporating Google Community Mobility Reports**

AUTHORS:

Nao Yamamoto[1], MPH

Haiyan Wang[2*], PhD

AFFILIATIONS:

[1]School of Human Evolution and Social Change, Arizona State University, Tempe, AZ 85287, USA.

[2]School of Mathematical and Natural Sciences, Arizona State University, Phoenix, AZ 85069, USA.

* Correspondence to:

Dr. Haiyan Wang. School of Mathematical and Natural Sciences, Arizona State University, Phoenix, AZ 85069, USA.

haiyan.wang@asu.edu, 602 543 5635

**Abstract (170/200)**

In June 2020, Arizona, U.S., emerged as one of the world's worst coronavirus disease 2019 (COVID-19) spots after the stay-at-home order was lifted in the middle of May. However, with the decisions to reimpose restrictions, the number of COVID-19 cases has been declining, and Arizona is considered to be a good model in slowing the epidemic. In this paper, we aimed to examine the COVID-19 situation in Arizona and assess the impact of human mobility change. We constructed the mobility integrated metapopulation susceptible-infectious-removed model and fitted to publicly available datasets on COVID-19 cases and mobility changes in Arizona. Our simulations showed that by reducing human mobility, the peak time was delayed, and the final size of the epidemic was decreased in all three regions. Our analysis suggests that rapid and effective decision making is crucial to control human mobility and, therefore, COVID-19 epidemics. Until a vaccine is available, reimplementations of mobility restrictions in response to the increase of new COVID-19 cases might need to be considered in Arizona and beyond.

**Keywords**

COVID-19; mobility restriction; mobility data; transmission model; simulation

## Background

In June 2020, Arizona, U.S., emerged as one of the world's worst coronavirus disease 2019 (COVID-19) spots with an alarming resurgence of COVID-19 after the stay-at-home order was lifted in the middle of May.[1,2] On June 29, Arizona reported 5,484 new COVID-19 cases, an all-time high record, averaged 3,680 daily new cases reported at the 27th week of 2020. However, with the decisions to reimpose restrictions such as closing bars, gyms, and some businesses, that average has been monotonically declining week-over-week, and on August 20, the average number of new cases in the past seven days reached 779. While many other states in the U.S. are experiencing the increase of new cases, Arizona is considered to be a suitable model in slowing the epidemic.[3]

In the absence of pharmaceutical control measures, mobility restriction is one of the few known interventions to mitigate the disease spread effectively.[4,5,6] However, level and type of restrictions have been questioned, and there is debate over when restrictions should be imposed or lifted.[7,8] The studies used to inform decision-makers are mainly based on parameter values obtained from other scenarios or countries, and it may not be suitable for the specific region. To aid in assessing the effect of mobility restrictions in Arizona, the use of daily aggregated real-world data on human mobility can be a helpful tool.[9,10,11,12]

In the present paper, we aimed to examine the COVID-19 situation in Arizona and assess the impact of change in human mobility. Combining the metapopulation susceptible-infected-removed (SIR) model of COVID-19 transmission with a dataset for COVID-19 cases and Google Community Mobility Reports (GCMR), we estimated how transmission differed in three regions in Arizona. Using estimated epidemiological parameters, we simulated how human mobility affects the epidemic of COVID-19. Our findings may provide useful insight when implementing a policy to mitigate the epidemic.

# Methods

## Data sources and collection

To estimate the COVID-19 transmission dynamics in Arizona, we fitted the mobility integrated metapopulation SIR model to publicly available datasets on Arizona cases. We compute the COVID-19 prevalence of each region by adding the cases of all counties belonging to a region. We use the COVID-19 data from the New York Times at the state and the county level over time. The New York Times compiles the time series data from state and local governments and health departments to provide a complete record of the ongoing outbreak.[13] Daily epicurve stratified by the data, and the timing of policy implementations are shown in Figure 1. Besides, we incorporate the GCMR to model the inter-region human movement. These reports provide insights into how people's social behaviors have been changing in response to policies aimed at combating COVID-19. The reports provide the changes in movement trends compared to baselines overtime at the U.S. county level, across different categories of activities: retail and recreation, groceries and pharmacies, parks, transit stations, workplaces, and residential.[14] Then we compute each region's daily changes by adding the changes of all counties belonging to a region. For the intra-region movement, we used Census commuting flows[15] and adjusted by the proportion of the population working from home estimated at 50%[16] (Table 1).

## Clustering counties

There are 15 counties in the U.S. state of Arizona. In general, Arizona is divided into Northern, Central, and Southern Arizona. Although the boundaries of the regions are not well-defined, as we aim to assess human mobility during the COVID-19 epidemic in Arizona, we group the counties of Arizona into the three regions in Figure 2. The three regions represent different geographical and social characteristics in terms of the spread of COVID-19. Much of the Northern region is National Forest Service land, parkland, and includes Navajo reservations. Central Arizona (Maricopa County), where the city of Phoenix is located, has about 60% of the population in Arizona, and more than half of COVID-19 cases are reported in this region. Southern Arizona includes Tucson and several other small cities close to the border of the U.S. and Mexico. In the following, we will use NAZ, CAZ, and SAZ to denote the clusters of the Arizona counties.

## Model structure and parameter estimation

We divided the population according to the infection status into susceptible (S), infected (I), and removed (R) individuals, and according to three sub-populations $i$. Susceptible individuals get infected at a given rate when they come in effective contact with an infectious individual, acquire infectiousness, and later either recover or die. We assumed Arizona to be a closed system with a constant population size of 7·3 million (i.e., S + I + R=7·3 million) throughout this epidemic. Additionally, we incorporated the effect of human mobility change based on GCMR. For a given region $i$, the evolution of the population in each status over time can be described by

$$\frac{dS_i}{dt} = -\left(\sum_{j=1}^{3} \beta_j \frac{I_j}{N_j} p_{ij} \phi_j(t)\right) S_i$$

$$\frac{dI_i}{dt} = \left(\sum_{j=1}^{3} \beta_j \frac{I_j}{N_j} p_{ij} \phi_j(t)\right) S_i - \gamma_i \widehat{\phi_i(t)} I_i$$

where $\beta_i$ is the transmission rate, $\gamma_i$ is the recovery rate, $p_{ij}$ describe the intra-region relative contact rate between region $i$ and region $j$ (i.e., $p_{ij} = 1$ for $i = j$). $\phi_i(t)$ is a function which describes activities outside of the home and $\widehat{\phi_i(t)}$ describes stay-at-home activities and it is written in a form

$$\phi_i(t) = \begin{cases} \frac{1}{1+\exp(-m_i(t-10))}, & i = j \\ 1, & otherwise \end{cases},$$

where $m_i(t-10)$ is empirical data for human mobility. We assume that time lags between mobility and COVID-19 cases are ten days based on our analysis (appendix p3).

The parameters for SIR were found using least-squares analysis and the Nelder-Mead optimization algorithm with 3,000 iterations. Models were fit to empirical infected cases per day.

**Estimating the impact of the timing of decision making and the final epidemic size**

Using parameters from the parameter estimation, we estimated the number of excess cases due to the delay in decision making using ten days as the baseline lag. Delays considered are one to 14 days, and LOESS curve fitting was used to fit excess cases compared to the baseline cases. We also examined the impact of human mobility on epidemic peak time by varying reduction rates 0%, 20%, and 40% as well as the real date from GCMR. For the real data, we assumed the mobility rate stays at the same level as the last day of available data (i.e., August 13, 2020). We estimated the final epidemic size for the mobility reduction rate from 0% to 70%.

We conducted a Pearson correlation analysis (appendix p4) to determine which activity is most closely related to COVID-19 cases. It is found that the retail and recreation category has the highest coefficients in general. Based on the result, we simulated how mobility reduction targeting retail and recreation change the disease trajectory and the final epidemic size.

**Role of the funding source**

## Results

For the COVID-19 epidemic in three Arizona regions, we estimated the model parameters summarized in table 1 based on data from March 1, 2020, to July 25, 2020. This period includes the most recent restrictions enforced since June 29, 2020, which have slowed down epidemic's spread. Figure 3 illustrates our estimates of COVID-19 cases in 3 regions during the abovementioned period. The red dotted lines represent the prevalence based on the observed COVID-19 cases, and blue lines represent the predicted COVID-19 cases. Our estimations underestimate the number of cases before the exponential growth phase; however, it captures the trend well after the phase. Here our model assumed that the mobility change is perceptible after ten days, which is estimated by our correlation analysis in the appendix (p3).

The relationship between mobility change over time for five categories of activities and state order timings is illustrated in the appendix (p2). Human mobility was significantly decreased after the decisions to reimpose the restrictions to close indoor bars, gyms, water parks, and movie theaters. We have examined excess cases due to the delay in policy implementation that varied over three regions, NAZ, CAZ, and SAZ (figure 4). The number of excess cases increased linearly for the first seven days of delay, while the prolonged delay in time did not produce more cases after the seven days. If the policy implementation falls behind by a week, NAZ, CAZ, and SAZ might have experienced at most 19,358, 157,467, 46,807 more cases, respectively.

Our simulations showed that by reducing human mobility, the peak time was delayed in all three regions. The peak time using the daily empirical data was similar to the mobility reduction rates of 20% (figure 5A). Specifically, a 20% reduction in mobility could prevent a peak occurring for a month, 40% could delay the peak for three more months. If we decrease 20% of mobility, we could reduce about 23% to 35% of cumulative cases at the end of the epidemic (figure 5B). For CAZ and SAZ, the cumulative incidences for empirical data were also similar to what we see in the case of a 20% reduction in mobility; however, cumulative incidence for NAZ resulted in the level of 0% mobility reduction. Figure 5C illustrates the final epidemic sizes in three regions based on the different reduction rates from 0% to 70%. With an overall 41% of mobility reduction, we can contain the virus resulting from reproduction number less than one. The mobility reduction is most effective for CAZ with more than double effect than for SAZ by an increasing 1% reduction rate.

We repeated the simulation targeting only retail and recreation, which showed the highest correlation with the COVID-19 growth ratio. Mobility restrictions on only retail and recreation could delay the epidemic's peak up to a month (figure 6A). By closing all the retail and recreation, the final epidemic size was reduced by about 35%. Although targeting retail and recreation did not decrease the magnitude as much as the scenario reducing all the activities at the same level, the peak time for full retail and recreation closure occurred around the same time with a 20% reduction in all the activities.

## Discussion

Strict control measures, such as stay-at-home orders, have been implemented to contain the COVID-19 outbreak in many countries, and it has been shown to be effective.[17,18,19,20] To evaluate the effect of policy implementation and its issued timing on the dynamics of COVID-19 spread, we incorporated empirical human mobility data in our metapopulation SIR model. We estimated parameter values, simulated outbreaks using the parameters, and assessed the impacts of mobility restrictions. Our findings suggest that rapid and effective policy implementation might have successfully slowed down the COVID-19 outbreak in Arizona.

Applications of digital technologies in public health have caught attention in the past few years[21] and it has been applied to assess this current COVID-19 pandemic.[9,22] Because mobility restrictions are known to have a significant role in mitigating the COVID-19 epidemic, several mathematical modeling studies incorporated human mobility data.[11,12] However, evidence of the direct effect of policy implementations on behavioral changes, which leads to mitigation of COVID-19 outbreak are scarce. Because our analysis is based on real-time data, we captured the effects of policy implementation directly without relying on assumed characteristics of COVID-19 and interventions' effectiveness.

The effect of mobility restrictions is known to be delayed a few weeks because of the incubation period of severe acute respiratory syndrome coronavirus 2 and reporting delay.[23,24] To estimate the parameter which describes the current COVID-19 epidemic in Arizona, we first estimated the time lag using Pearson correlation analysis. Our analysis revealed that the impact of mobility changes on COVID-19 cases is observed ten days after the policy shift, which agrees with what Badr et al.[9] estimated. Unlike other studies, we were able to analyze not only the decrease in COVID-19 cases due to policy implementation but also the increase in cases due to a lifting of stay-at-home order followed by another decrease. Our study focuses on Arizona because Arizona, previously recorded as one of the worst hot spots of the COVID-19 epidemic in the world because of the early reopening of the economy, started to decline significantly in cases after the state reimposed restrictions.

Geographic heterogeneity in Arizona should also be mentioned. We divided Arizona into NAZ, CAZ, and SAZ because they represent different geographical and social characteristics regarding the spread of COVID-19. NAZ is considered to be rural; it is sufficient to prevent the cluster of cases in places like a gym or bar. CAZ is the metro city where more than 60% of the COVID-19 cases are reported. Because the population size and epidemic size are large, it is not enough to target only retail and recreation, but it is vital to reduce mobility overall. Since the epidemic size can quickly blow up, the policy implementation's timing is also crucial. SAZ includes the city of Tucson and several other small cities close to the border of the U.S. and Mexico, containing the disease is more complex despite their population size. Therefore, we have simulated three regions separately to capture the actual dynamics.

In the basic scenario with empirical data, the epidemic peak time in NAZ was predicted in late September, and CAZ and

SAZ have already peaked in early July. Simulation of 0%-70% mobility reduction in each region revealed that mobility reduction exceeds 41% is not necessary to mitigate COVID-19 spread. While mobility restrictions are crucial to containing the virus, there is severe debate over the level of restrictions considering economic impacts and other possible harms attributed to school closure.[25,26] In this regard, we have conducted another Pearson correlation analysis and discovered that retail and recreation have the strongest correlation with the COVID-19 growth rate. We examined the epidemic dynamics with mobility reductions targeting only retail and recreation. This intervention was most effective in NAZ, preventing 32% of the population from getting infected by closing all the retail and recreation. For CAZ and SAZ, the intervention to only target retail and recreation was not sufficient to control the disease spread because of high density and transmissibility.

Our study has several limitations. First, because there was no available data on recovered cases, we estimated using criteria developed by the Texas Department of State Health Services (appendix p1). The calculations based on the estimates of fraction hospitalization in China may alter the estimated recovered cases. Second, intra-county mobility fractions were based on Census commuter data from 2015 that might not necessarily reflect mobility flow during the COVID-19 epidemic. However, the fraction relative to inter-county mobility is unlikely to change due to the epidemic; we believe we reasonably estimated the parameters by scaling with a work-from-home proportion of 50%.

## Conclusions

In conclusion, we have constructed the metapopulation SIR model and fitted two sets of publicly available data to assess the effect of policy implementation. While the situation around COVID-19 in Arizona has been improving, the epidemic is not yet under control, and a large proportion of the population is still susceptible. As the trajectory of the outbreak in Arizona and beyond will depend on human mobility, implementing rapid and effective control measures is advised based on Arizona's experiences.

**Funding**

This wark was partially supported by the National Natural Science Foundation (#1737861); Nishihara Cultural Foundation.

**Figure legends**

**Figure 1 | Geographical distribution and temporal dynamics of confirmed COVID-19 cases**

Time series of COVID-19 cases and policy shift in Arizona.

**Figure 2 | Three regions in Arizona**

We divided 15 counties in the U.S. state of Arizona into Northern, Central, and Southern Arizona.

**Figure 3 | Estimate of COVID-19 cases in three regions in Arizona**

Epidemic evolution predicted by the metapopulation SIR model using confirmed COVID-19 cases from NAZ, CAZ, and SAZ during March 1, 2020, and July 25, 2020. The red dotted lines represent the prevalence based on the confirmed COVID-19 cases, and blue lines represent the predicted COVID-19 cases.

**Figure 4 | Excess cases due to delay in policy implementation compared to the baseline scenario**

The blue lines represent the mean number of extra cases that might occur due to the delay in policy implementation, and the shaded area represents 95% confidence intervals.

**Figure 5 | Effects of human mobility in Arizona on COVID-19 cases**

Daily prevalence of COVID-19 (A) and cumulative incidence of COVID-19 (B) with mobility restrictions varied 0%, 20%, and 40%, as well as empirical data on mobility change. (C) The final size of the epidemic with varying mobility reductions ranged from 0% to 70%.

**Figure 6 | Effects of mobility reductions in retail and recreation settings in Arizona on COVID-19 cases**

Daily prevalence of COVID-19 (A) and cumulative incidence of COVID-19 (B) with mobility restrictions in retail and recreation settings varied from 0% to 100%.

Table 1. Parameter values used for SIR model for three regions

| Region | Population | $p_{ij}$: From NAZ to | $p_{ij}$: From CAZ to | $p_{ij}$: From SAZ to |
|---|---|---|---|---|
| NAZ | 848,693 | 1 | 0·0012 | 0·0018 |
| CAZ | 4,485,414 | 0·013 | 1 | 0·049 |
| SAZ | 1,944,610 | 0·0025 | 0·005 | 1 |

Table 2. Estimated parameter values for the metapopulation SIR model

| Parameters | $\beta_1$ | $\beta_2$ | $\beta_3$ | $\gamma_1$ | $\gamma_2$ | $\gamma_3$ |
|---|---|---|---|---|---|---|
| Values | 0·299 | 0·867 | 1·146 | 0·198 | 0·514 | 0·744 |

**Figure 1**

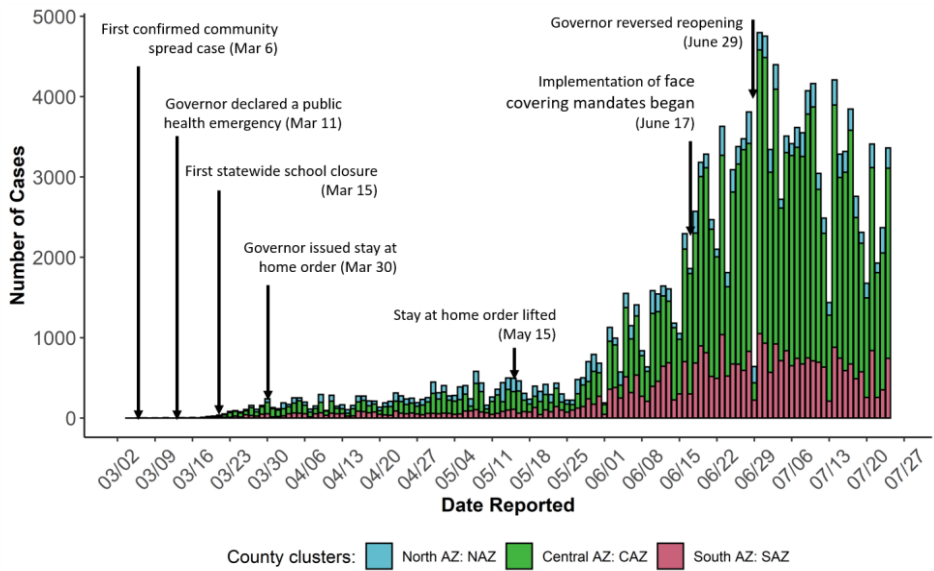

**Figure 2**

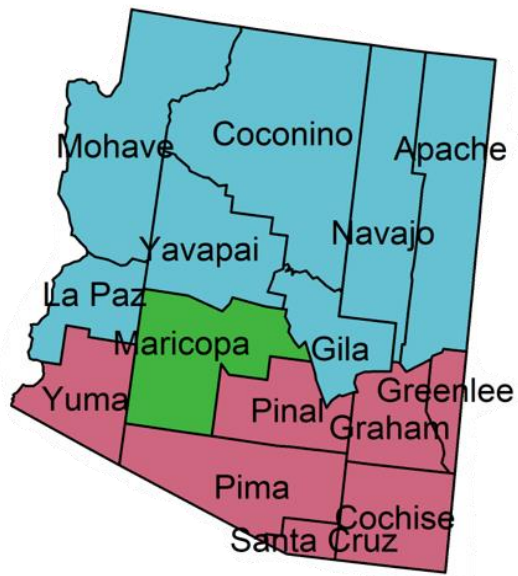

# Figure 3

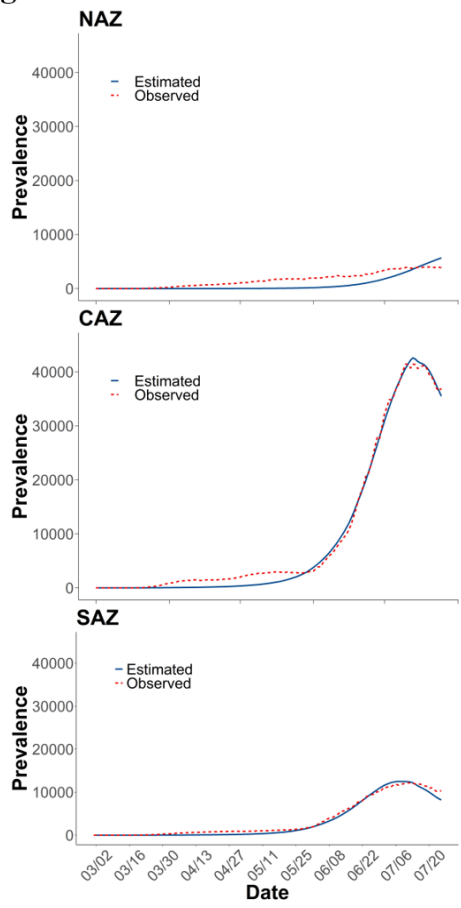

# Figure 4

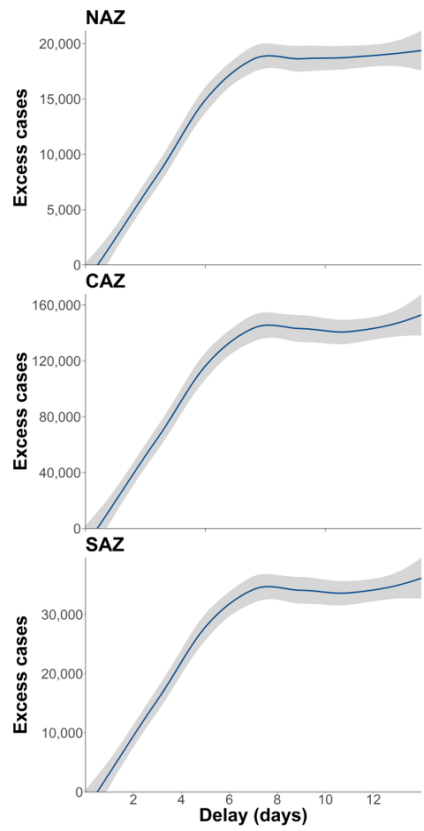

**Figure 5**

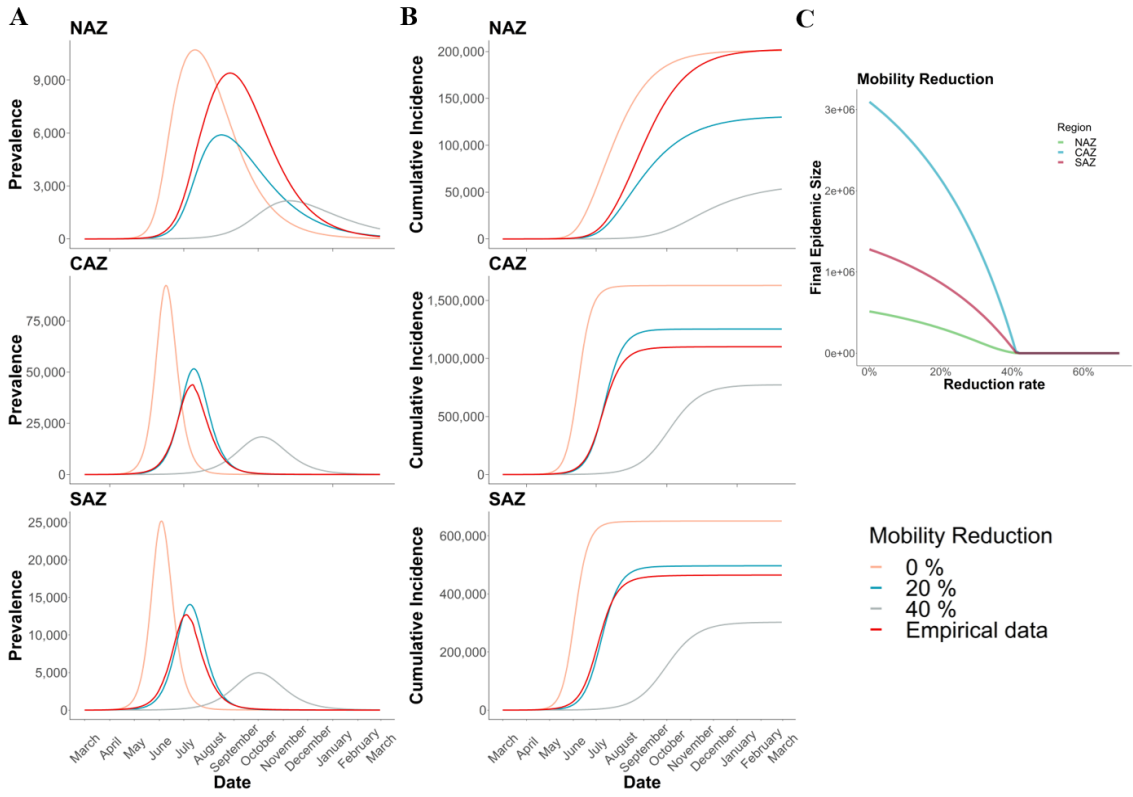

**Figure 6**

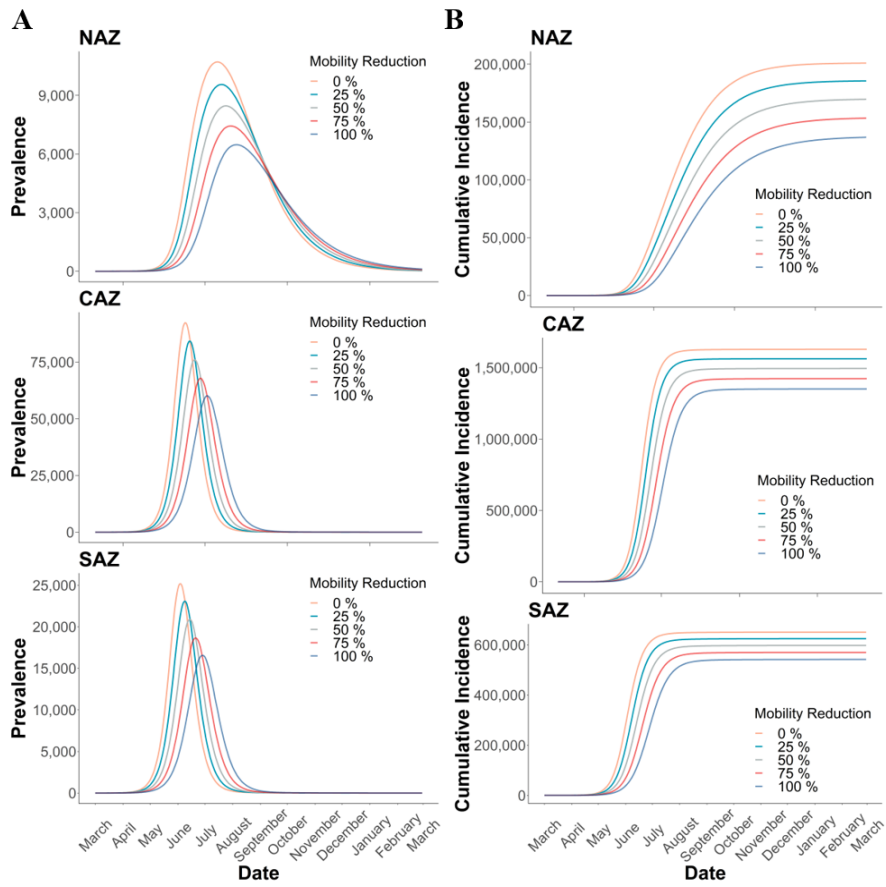

Supplementary Materials for

# Assess the impacts of human mobility change on COVID-19 dynamics in Arizona, U.S.: a modeling study incorporating Google Community Mobility Reports

Nao Yamamoto[1], MPH, Haiyan Wang[2*], PhD

[1]School of Human Evolution and Social Change, Arizona State University, Tempe, AZ 85287, USA.

[2]School of Mathematical and Natural Sciences, Arizona State University, Phoenix, AZ 85069, USA.

* Correspondence to: haiyan.wang@asu.edu

## Appendix Contents

**Supplementary Text**

Texas Department of State Health Services calculates the number of recovered cases by:

$$R(t) = (T(t-14) - D(t-14)) * 0.8 + (T(t-32) - D(t-32)) * 0.2$$

where $T(t)$ is the total number of cases, $D(t)$ is the total number of death cases. Here we are assuming that 20% of cases required hospitalization, and hospitalized patients require about 32 days to be recovered while 80% did not receive hospitalization and recovered after 14 days.[1,2]

**Figure S1**

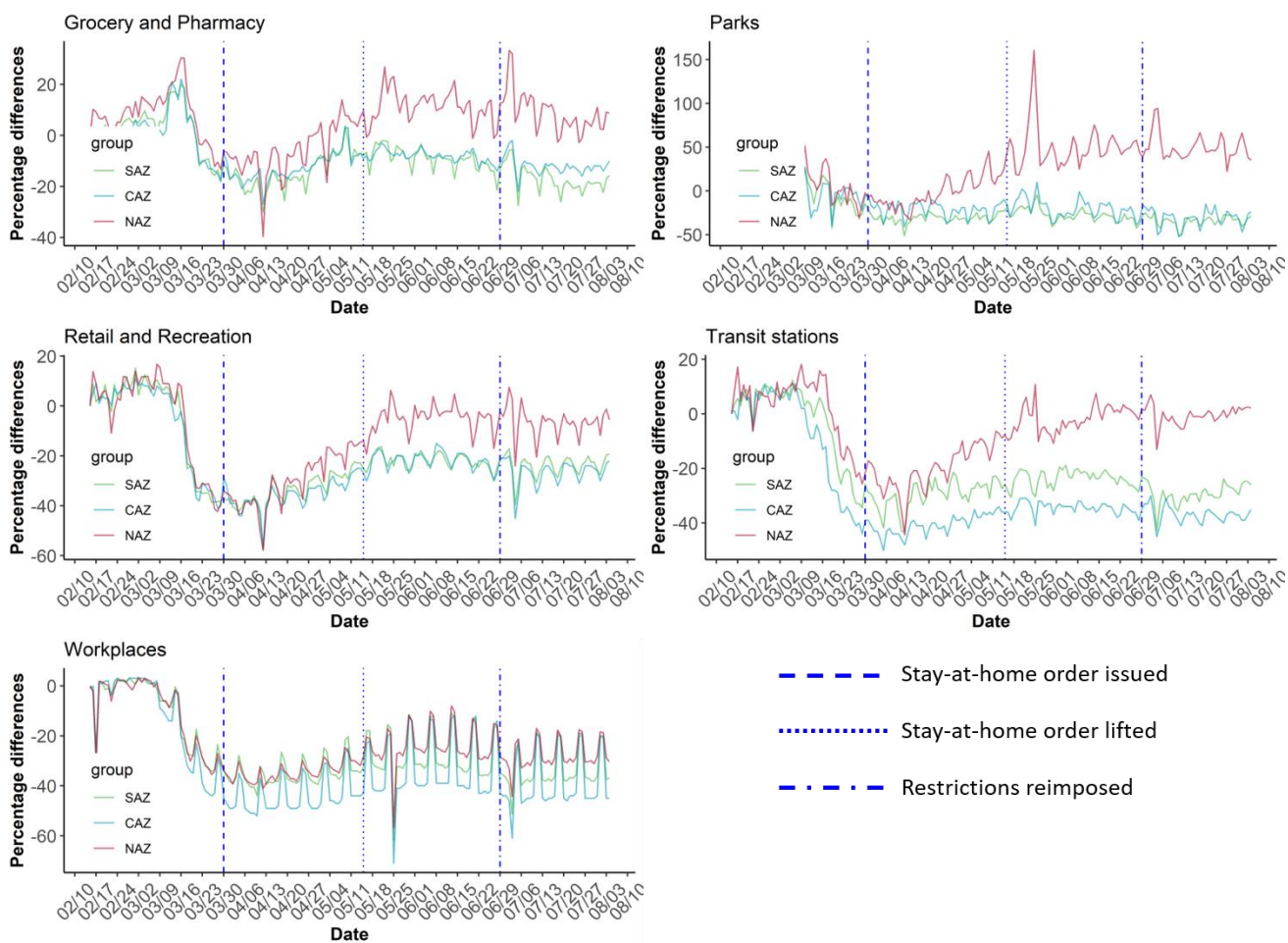

**Mobility change data in Arizona retrieved from Google Community Mobility Reports.**

Mobility trends over time across different categories of places such as retail and recreation, groceries and pharmacies, parks, transit stations, and workplaces in three regions are shown.

Blue dashed lines represent the day that stay-at-home order was implemented in Arizona (2020-03-30), blue dotted lines are for the day stay-at-home order was lifted (2020-03-15), and blue dot-dashed lines represent the day Governor reimposed the restriction (2020-06-29).

**Figure S2**

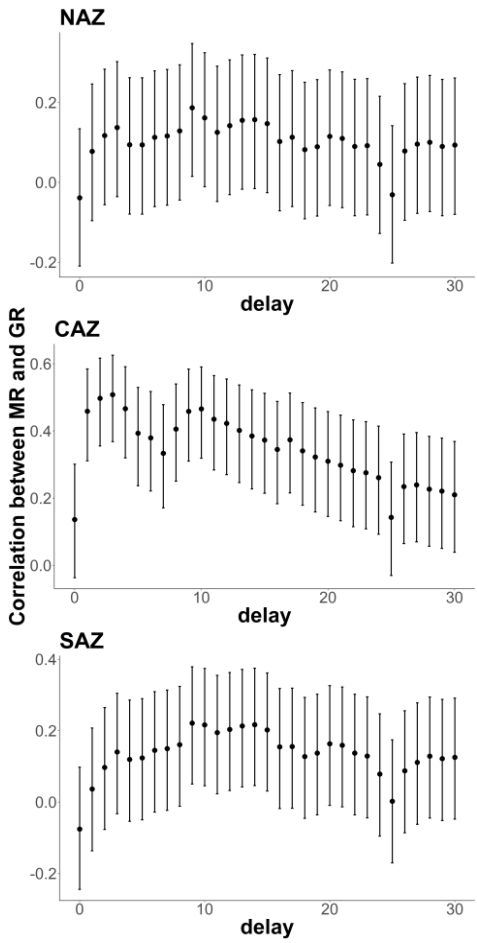

Correlations between mobility ratio (M.R.) and growth rate ratio (G.R.) of COVID-19 cases at different lags (in days) for each region.

Following Badr et al.,[3] the mobility ratio was obtained from Google Community Mobility Reports, which represents the percentage difference compared to the baseline. The growth rate ratio was calculated as follows:

$$GR = \frac{\log\left(\sum_{t-2}^{t} \frac{C(t)}{3}\right)}{\log\left(\sum_{t-6}^{t} \frac{C(t)}{7}\right)}$$

where $C(t)$ is the number of COVID-19 cases on day $t$.

**Figure S3**

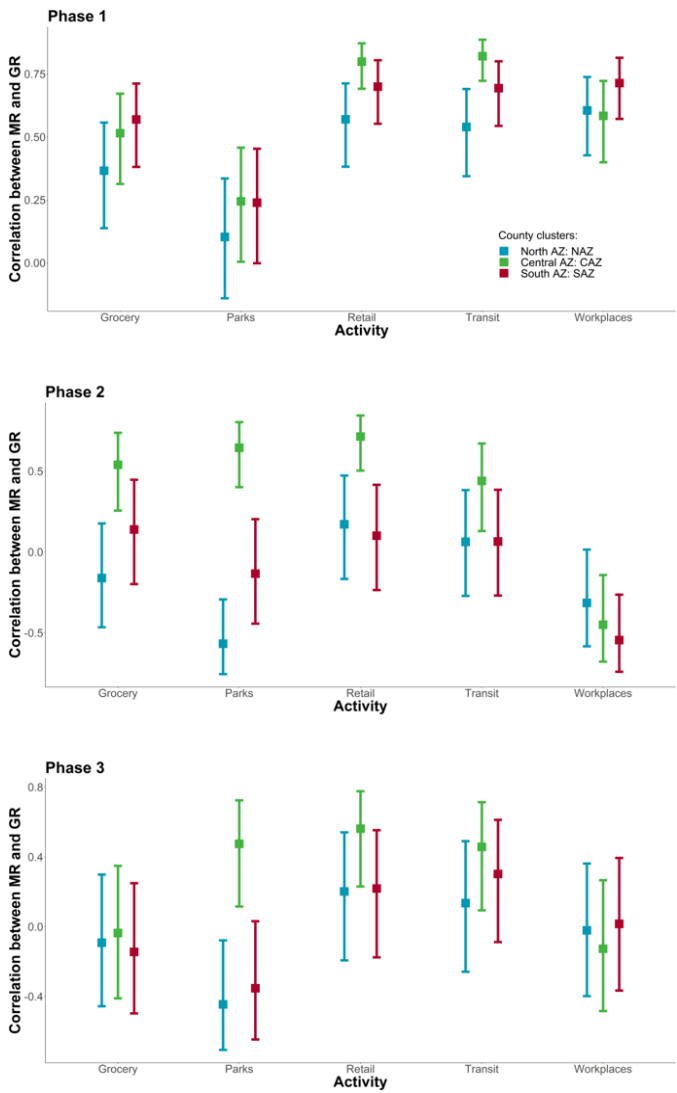

**Correlation analysis of human mobility and COVID-19 growth rate for three time periods.**

(A) Phase 1: During stay-at-home order (2020/3/30 – 2020/5/15 ).

(B) Phase 2: From the day stay-at-home order was lifted to the day which Governor reimposes the restrictions (2020/5/15 – 2020/6/29).

(C) Phase 3: From the reimpose to the present (2020/6/29 – 2020/8/04).

**Appendix references**